\documentstyle[twocolumn,prl,aps]{revtex}

\def\beqn{\begin{eqnarray}}
\def\eeqn{\end{eqnarray}}
\def\beqns{\begin{eqnarray*}}
\def\eeqns{\end{eqnarray*}}
\def\beq{\begin{equation}}
\def\eeq{\end{equation}}
\def\bea{\begin{array}}
\def\ea{\end{array}}
\def\one{1\hskip-.37em 1}
\def\<{\langle}
\def\>{\rangle}

\draft
\widetext

\begin{document}

\twocolumn[\hsize\textwidth\columnwidth\hsize\csname
 @twocolumnfalse\endcsname
\draft
\preprint{}

\title{\sc A new formalism for the computation of RKKY interaction 
in aperiodic systems}

\author{\sc Stephan Roche ${\ }^{\dagger}$ and Didier Mayou${\ }^{*}$}
\address{${\ }^{\dagger}$ Department of Applied Physics, University of Tokyo, 7-3-1 Hongo, Bunkyo-ku, Tokyo 113, Japan.\\
${\ }^{*}$ LEPES-CNRS, avenue des Martyrs- BP166 38042 Grenoble France}

\maketitle

\begin{abstract}
\leftskip 54.8pt
\rightskip 54.8pt
A numerical method to investigate RKKY interaction between localized
spins in aperiodic materials is presented. Based on an expansion of the spectral measure on orthogonal polynomials, this approach leads to an effective framework to analyse linear response formula for the RKKY coupling in complex systems. 
Particularly useful in the tight-binding scheme it is used in this paper to probe the role of the local environment and the quasiperiodic potential on the interaction between magnetic spins. Interesting features are revealed and discussed within the context of anomalous localization and transport.
\end{abstract}


]

\vspace{20pt}

\section{Introduction}

\vspace{10pt}

\hspace{\parindent}The purpose of this work is to present a new formalism based on 
real space recursion scheme \cite{Magnus} which enables to compute the so-called 
Rudermann Kittel Kasuya Yosida interaction (RKKY)~\cite{RKKY,Degennes,Jag,Lerner} 
an effective coupling between two localized magnetic moments, mediated by conduction electrons.

\vspace{10pt}

\hspace{\parindent}This long range oscillatory interaction is now well understood in
 pure metals and in weakly disordered systems, or even close to the Anderson 
 transition. RKKY interaction has been the subjected of a great attention during the past years for instance for
understanding the spin glass transition \cite{Spinglass}, magnetic long range order
in high-$T_{c}$ cuprates\cite{HTC}, or more recently because of its anomalous 
behavior related to giant magnetoresistance effects in magnetic multilayers \cite{GMR}.
 
\vspace{10pt}

\hspace{\parindent}As conduction electrons carry this interaction, one can wonder
 what happens for systems in which electronic propagation is anomalous \cite{RochePRL}
  or in cases of inhomogeneous disorder ? (e.g. local inhomogeneities, correlated 
  disorder,...). These issues may be addressed in relation to unusual RKKY interaction 
  in quasicrystals, where the amplitude of the coupling was found to be anomalously 
  strong and with similar values for several manganese-based quasicrystals\cite{Berger}
  with different densities of states. Such results were found to apparently \
  contradict the classical theory stating that RKKY should be proportional to the total DOS at Fermi level.

\vspace{10pt}

\hspace{\parindent}In the following, after reminding some general 
features of RKKY coupling, the method will be detailed followed by a 
first part addressing the 
numerical convergence of the algorithm. A second part 
is devoted to the study of the impact of local disorder or quasiperiodic potential on RKKY. This will exemplify the interest of the method.

\section{Calculation of the non-local susceptibility}

\vspace{10pt}

\hspace{\parindent}Indirect RKKY interaction stems from the coupling between localized magnetic moments and propagating electrons. If an electron in a state of energy
 $E<E_{F}$ undergoes a transition to a state of energy
 $E'>E_{F}$ because of the coupling with the localized moment in $\mid r_{i}\>$, 
 then a hole-electron pair is created and propagate coherently during a 
 certain time $\tau$, with  $|E'-E| \leq \hbar/\tau$, until the pair is 
 destroyed by diffusion on another magnetic impurity located in  $\mid r_{j}\>$. 
 Consequently the longer is the propagation time the smallest will be the vicinity 
 around Fermi energy that account for RKKY. The generic form of the effective 
 coupling between two magnetic impurities mediated by itinerant electrons reads :

\vspace{5pt}

$$
{\cal I}_{RKKY}(r_{i},r_{j},E)=J^{2}\chi(r_{i},r_{j},E)
 {\bf S}_{r_{i}}.{\bf S}_{r_{j}}$$

\noindent
with J is the interaction between
the localized moment ${\bf S}_{r_{i}}$  and the spin of the itinerant
electrons, and $\chi(r_{i},r_{j},E_{F})$
contains the sum of all the electron-hole propagation paths from
 $|r_{i}\>$ to $|r_{j}\>$. RKKY is then proportional to the electronic susceptibility $\chi (r_{ij})$ of itinerant electrons. When $J>0$ (resp. $J<0$), the configurations of parallel spins 
(resp. antiparallel) will minimize the energy promoting ferromagnetic state (resp. antiferromagnetic state). The susceptibility as a contribution of all the scattering pathes of the hole-electron pair can
be written down as

$$
\chi(r_{i},r_{j})=-\frac{1}{2\pi}\Im m\int_{-\infty}^{+\infty}dE \
 G_{+}(r_{i},r_{j},E)G_{-}(r_{j},r_{i},E)
$$

\noindent
introducing the retarded ($G_{-}$) and advanced Green's functions ($G_{+}$) which define the 
amplitude of propagation of the hole-electron pair. Note that there exists 
an exact sum rule between the susceptibility and the local density of states (LDoS)
 $\sum_{j}\chi(r_{i},r_{j})=\rho_{i}(E)= {-2\over \pi} \Im m\<r_{i}\mid G(z)\mid r_{i}\>$ 
 which could be used as a numerical test.

\vspace{5pt}

\hspace{\parindent}
In metallic systems with space dimension D, the interaction is given by

\begin{equation}
{\cal I}_{RKKY}(r,E_{F})\sim A({\bf r})\cos(2k_{F}r+\delta({\bf
r}))/r^{D}
\end{equation}

\noindent
which manifests a long range oscillating behavior.
 \cite{RKKY} For a free electron gas, the  $A, \delta$ are independent of $ {\bf r}$, 
 whereas for weak disorder limit, $A({\bf r})$ becomes a random but smooth function of
  ${\bf r}$ and $\delta({\bf r})$ is the phase shift associated to the scattering 
  of electrons on impurities. $\delta({\bf r})$ becomes random for 
 $r>l_{pm}$ (mean free path). It has
 been shown however that the mean free path is not a 
 good measure of the typical length scale of this interaction. 
 Indeed, the disorder average of even moments {
 \small $\<\chi^{2p}(\mid r_{i}-r_{j}\mid)\>$} 
 will contribute roughly as $(1/r^{D})^{1/2p}$, 
 whereas odd moments of the susceptibility will vanish exponentially 
 over the mean free path length as expressed by the following averaged results \cite{Lerner}

{\small
\begin{eqnarray}
\<\chi^{2p}(\mid r_{i}-r_{j}\mid)\>\ &\simeq& \ \Omega_{p}\biggl 
( {{\rho^{2}(E_{F})}\over{\mid r_{i}-r_{j}\mid^{2d}}}\biggr)^{p}\sim 
\bigl(\<\chi^{2}(\mid r_{i}-r_{j}\mid)\>\bigr)^{p}\nonumber  \\
\<\chi^{2p+1}(\mid r_{i}-r_{j}\mid)\>\ &\simeq&  \ 
\hbox{exp}(-\frac{\mid r_{i}-r_{j}\mid}{\strut l_{pm}}) \nonumber
\end{eqnarray}
}

\noindent
taking $\rho(E_{F})$ the DoS at Fermi level, $l_{m}$ the mean free
path, and $\Omega_{p}$ a constant independent of the parameter of 
the hamiltonian. Consequently, the average over disorder of electronic
susceptibility is not sufficient for describing the correct range, 
amplitude and phase of the interaction, which fluctuates very much from one 
random configuration to another.~\cite{Jag,Lerner}

\vspace{5pt}

\hspace{\parindent}Let us now consider a simple argument for the RKKY coupling. 
If one considers a cube of length $L$ (with periodic boundary conditions) which is larger
 than the typical distance magnetic impurities, then the expansion of the susceptibility
  in a basis of eigenstates  $\{|n\>\}$, reads :

\vspace{5pt}

\beq
\chi(r_{i},r_{j})=
2 \Re e  \sum_{n,m} \frac{\<r_{i}|n\>\<n|r_{j}\>\<r_{j}|m\>\<m|r_{i}\>}{E_{n}-E_{m}}
\eeq
\vspace{5pt}

\noindent
with {\small $E_{n}>E_{F}>E_{m}$} and  {\small $\<n|r_{j}\>\sim 1/\sqrt{L^{D}}$} given 
that states are normalized. Given that the average spacing between energy levels for a
 given length is roughly {\small $\Delta E=(\rho(E_{F})L^{D})^{-1}$},
with $\rho(E)$ the total density of states,
  one can assume that the energies of the electron and hole are respectively 
  {\small $E_{n}=E_{F}+n\Delta E$ } and  {\small $E_{m}=E_{F}-m\Delta E $} ($n,m>0$). 
  Consequently, the susceptibility can be expressed as

\vspace{5pt}

{\small
\begin{eqnarray}
\chi(r_{i},r_{j})&=&
2  (L^{-D})^{2}\ \Re e  \sum_{n,m}\ \frac{\Phi(n)\Phi^{*}(m)}{(n+m)\Delta E}
\nonumber\\
&=& \frac{2\rho(E_{F})}{L^{D}}\  \Re e 
\sum_{n,m}\ \ \frac{\Phi(n)\Phi^{*}(m)}{(n+m)}\nonumber
\end{eqnarray}
}

\noindent
where the functions $\Phi(n)\sim L^{D} \<r|n\>\<n|r'\>\sim 1$. From this, one can extract the generic behavior in power laws  $L^{-D}$, which only depends on the space dimension in which the system is embedded. The other part accounts for phase interferences and for instance in the case of almost free electrons (weak disorder limit), {\small $\Phi(n)=\exp(i{\bf k}.(r_{i}-r_{j}))$} so that 
{\small $\Re e\Phi(n)\Phi^{*}(m)\sim e^{-ik_{F}(r_{i}-r_{j})}
e^{+ik_{F}(r_{i}-r_{j})}\sim \cos(2k_{F}|r_{i}-r_{j}|)$} given that only states (m,n) very
 close to the Fermi surface will contribute to the electronic response. Possible limitations
 of this treatment due to strong correlations between electronic levels will 
 be discussed later on (see part V.C).

\vspace{5pt}

\hspace{\parindent}From our argument, two different effects influencing the physical 
properties may be drawn. On the one hand, close to metal-insulator transition 
peculiar electronic localization (like multifractality of eigenstates) 
may lead to correction of the law $1/|r_{i}-r_{j}|^{D}$. Indeed, in quasiperiodic 
systems, quasicrystals or disordered fractal structures, the average displacement 
in time of an electron is driven by anomalous diffusion. One can wonders how 
electronic susceptibility is affected by such localization effects.

\vspace{5pt}

\hspace{\parindent}Besides, if the LDoS is inhomogeneous, i.e if it presents strong local 
fluctuations, an effect of local environment
may be deduced from the equation (2) and lead to specific features for RKKY coupling. 
Actually, since the LDoS $\mid r\>$ can be expanded in the eigenstates 
basis of ${\cal H}$ as {\small $\sum_{n} \delta(E-\varepsilon_{n})\mid 
\< r\mid n\>\mid^{2}$}, a substantial increasing of the LDoS around the magnetic sites,
enhancing the corresponding amplitudes $\< r\mid n\>$, should also qualitatively
lead to an increase of the non-local susceptibility. This has been discussed 
in relation with peculiar properties of AlPdMn quasicrystalline phases 
where the RKKY interaction was found to be anomalously large and 
identical for several phases with different TDoS \cite{Berger,Roche,RO-RKKY}.

\vspace{5pt}

\section{Real-space approach of RKKY interaction}

\vspace{10pt}

\hspace{\parindent}
To perform real space calculations of the non-local susceptibility, considering the Green operator {\small $G(z)=(z-{\cal H})^{-1}=\int_{-\infty}^{+\infty}
\frac{\delta (E-{\cal H})}{ z-E} dE$}, one starts from the general form 
of $\chi(r_{i},r_{j})$ :

{\small
$$2 \Re e \int_{E>E_{F}\atop
E'<E_{F}}dEdE' {{\<r_{i}\mid \delta (E-{\cal H})\mid r_{j}\>\<r_{j}
\mid \delta(E'-{\cal H})\mid
r_{i}\>}\over{E-E'}}$$
}

\hspace{\parindent}The aim of the method is to determine the coefficients
$\<r_{i}\mid \delta(E-{\cal H})\mid r_{j}\>$ without exact diagonalization, usually limited to simple models and small finite size systems. The keypoint of the method 
is to use a basis of orthogonal polynomials $\{{\cal P}_{n}(E)\}_{n\in N}$ 
associated to a normalized function $\rho(E)$, referred as 
a model density of states. If the spectral subset of $\rho(E)$ contains
the one of the real hamiltonian, it can be shown that\cite{Roche}

$$\delta(E-{\cal H})=\rho(E)\sum_{n} {\cal P}_{n}(E){\cal P}_{n}({\cal H})$$

\noindent
and the $\{{\cal P}_{n}(E)\}_{n\in N}$ satisfy to the orthogonality 
condition

$$\int_{-\infty}^{+\infty}
\rho(E){\cal P}_{n}(E){\cal P}_{m}(E)dE=\delta_{nm}$$

\vspace{5pt}

\noindent
These relations enable to write {\small $\<r_{i}\mid \delta(E-{\cal H})\mid r_{j}\>= \rho(E)\sum_{n\in N}{\cal P}_{n}(E)\alpha_{ij}^{n}$} with 
$\alpha_{ij}^{n}=\<r_{i}\mid {\cal P}_{n}({\cal H})\mid r_{j}\>$.
 From these expressions, 
the susceptibility can be written down

\begin{eqnarray}
\chi(r_{i},r_{j})&=&\Re e  \sum_{nm} {\cal I}_{mn}\alpha_{ij}^{n}
\alpha_{ji}^{m} \nonumber\\
{\cal I}_{mn}&=&\int_{E>E_{F}\atop E'<E_{F}} \rho(E)\rho(E')
{{{\cal P}_{m}(E){\cal P}_{n}(E')}\over{E-E'}}dEdE'
\nonumber 
\end{eqnarray}

\vspace{5pt}

\hspace{\parindent}Accordingly, the calculation of 
$\chi(r_{i},r_{j})$ is divided into 
two independent parts. Depending on the choice of polynomials, the 
${\cal I}_{mn}$ will display particular analytical form. 
The other part of the susceptibility will involve a recursive evaluation of 
coefficients $\alpha_{ij}^{n}$ by means of the three term relations defining the 
orthogonal polynomials, as detailed below.

\subsection{Recursion evaluation of $\alpha_{ij}^{n}$}

\vspace{5pt}

\hspace{\parindent}One 
has to choose a convenient basis of orthogonal polynomials to evaluate step by step
the coefficients $\alpha_{ij}^{n}$. Given that any basis of such polynomials
is uniquely define by a three-term recurrence relation, generically reading
$EP_{n}(E)=a_{n}{\cal P}_{n}(E)+b_{n}{\cal P}_{n+1}(E)+b_{n-1}{\cal P}_{n-1}(E)$
with $b_{-1}=0, n\geq 0$, and $a_{n}, b_{n}$ the associated recursion
 coefficients~\cite{Szego}, the main vectors to be evaluated recursively
  will follow from
 
 $$\mid \varphi_{i}^{n}\>={\cal P}_{n}({\cal H})\mid r_{i}\>=\sum_{j}
\alpha_{ji}^{n}\mid r_{i}\>$$

\vspace{5pt}

\hspace{\parindent}Pratically, we will consider the 
Chebyshev polynomials of second order that have been already used in others contexts. 
Such polynomials are defined by

$${\cal P}_{n}({\cal H})=\frac{1}{b}({\cal H}-a)
{\cal P}_{n}({\cal H})+{\cal P}_{n-1}({\cal H})$$

\noindent
with ${\cal P}_{-1}({\cal H})=0$ and ${\cal P}_{0}({\cal H})=\one$ and the 
corresponding weight is given by

$$\rho_{ab}(E)=\frac{1}{2\pi b^{2}}\sqrt{4{b}^{2}-(E-a)^{2}}$$

\noindent
which is $\neq 0$ only for $E\in [a-2b,a+2b]$. The coefficients a and b are given
by the calculated limits  $a_{n\to\infty}, b_{n\to\infty}$ for the real densities of states. 
From the abovementionned relations, the $\mid \varphi_{i}^{n}\>$ will be given by 
 $\mid \varphi_{i}^{n}\>=\frac{1}{b}({\cal H}-a)\mid \varphi_{i}^{n}\>
 +\mid \varphi_{i}^{n-1}\>$ and $\mid \varphi_{i}^{-1}\>=0, 
 \mid \varphi_{i}^{0}\>=\mid r_{i}\>$. In the tight-binding scheme ${\cal H}=\sum_{pq} \gamma_{pq}
 \mid r_{p}\>\< r_{q}\mid$, one shows that the $\alpha_{ij}^{n}$ coefficients have to be evaluated recursively through ($\alpha_{pj}^{-1}=0, 
 \alpha_{pj}^{0}=\delta_{pj}, \forall p$)
 
 $$\alpha_{ij}^{n+1}=\frac{1}{b}(\sum_{p}\alpha_{ip}^{n}\gamma_{pj}-a
 \alpha_{ij}^{n})-\alpha_{ij}^{n-1}$$

\noindent

\vspace{5pt}

\subsection{Calculation of ${\cal I}_{mn}$ for the Chebyshev polynomials}

\vspace{10pt}

\hspace{\parindent}Let us now proceed to the calculation of ${\cal I}_{mn}$ 
for the Chebyshev polynomials of second order. First we rearrange the general form

$${\cal I}_{mn}=\int_{E>E_{F}\atop E'<E_{F}} \rho(E)\rho(E')
{{{\cal P}_{m}(E){\cal P}_{n}(E')}\over{E-E'}}dEdE'$$

\noindent
by noticing that the factor $1/(E-E')$ can be written as

{\small
\begin{eqnarray}
\oint_{\Gamma} \frac{dz}{(z-E)(z-E')}&=&\frac{2i\pi}{E-E'}
\bigl(\Theta(E'-E_{F})\Theta(E_{F}-E)\nonumber\\
&-&\Theta(E-E_{F})\Theta(E_{F}-E')\bigr)\nonumber
\end{eqnarray}
}

\noindent
with the Heaviside function ($\Theta(x)=0, x<0$ and $\Theta(x)=1, x>0$) and
the contour $\Gamma$ is shown in the complex plane on Fig.1 
for $\eta\to 0, R_{\Gamma}\to\infty$. One then rewrites :


{\small
\begin{eqnarray}
\Re e({\cal I}_{mn}+{\cal I}_{nm})&=&-\frac{i}{2\pi}\oint_{\Gamma}dz
\biggl(\int_{-\infty}^{\infty}dE\frac{\rho(E){\cal P}_{n}(E)}{z-E}\biggr)
\times  \nonumber\\
&\times& \biggl ( \int_{-\infty}^{\infty}dE'\frac{\rho(E'){\cal P}_{m}(E')}{z-E'}\biggr)
\nonumber
\end{eqnarray}
}

\vspace{5pt}

\hspace{\parindent}By application of Jordan Lemme, the integral 
on the contour $\Gamma$ tends to zero when the radius goes to infinity, and
it only remains four integrals on the real axis respectively for
 {\small $[-\infty,a-2b],\ [a-2b,E_{F}],\ [E_{F},a+2b],\ [a+2b,+\infty]$}. 
 Using the relation between first and second order Chebyshev polynomials defined on 
 $[-1,+1]$ and associated with $\rho(E)=\sqrt{1-E^{2}}$
 
 $$
\lim_{\eta\to 0^{\pm}}\int_{-1}^{+1}
{{\sqrt{1-E^2}P_{n}(a+2bE)}\over{\omega+i\eta-E}}dE=$$

$$\pi\{ Q_{n+1}(\omega)\mp i\pi\sqrt{1-{\omega}^2}P_{n}(\omega)\}
$$

\noindent
it is easy to show that for $\mid\omega\mid\leq 1$, with $\omega=\cos\phi$ 
then 

\begin{eqnarray}
Q_{n}(\omega)&=&\cos n\phi \nonumber\\
P_{n}(\omega)&=&\frac{\sin (n+1)\phi}{\sin\phi} \nonumber
\end{eqnarray}

\noindent
and finally

{\small
$$\lim_{\eta\to 0^{\pm}}\int_{-1}^{+1}
{{\sqrt{1-E^2}P_{n}(a+2bE)}\over{\omega+ i\eta-E}}dE=
\pi\exp(\mp i(n+1)\phi)$$}

\vspace{5pt}

\hspace{\parindent}In conclusion, given that the integration
 outside $[-1,1]$ leads to pure imaginary terms, the calculation of ${\cal I}_{mn}$ reduces to

{\small
\begin{eqnarray}
\Re e({\cal I}_{mn}+{\cal I}_{nm})&=&\frac{1}{2\pi b} \int_{\frac{E_{F}-a}{2b}}
^{1} d\omega\sin((m+n+2)\phi)\nonumber\\
&-&\frac{1}{2\pi b}\int_{-1}^{\frac{E_{F}-a}{2b}}
d\omega\sin((m+n+2)\phi)\nonumber\\
&=&\frac{1}{2\pi b}\biggl(\frac{\sin(m+n+3){\cal A}_{F}}{m+n+3}-
\frac{\sin(m+n+1){\cal A}_{F}}{m+n+1}\biggr)\nonumber
\end{eqnarray}
}

\noindent
where ${\cal A}_{F}=$Arcos$({{E_{F}-a}\over{2b}})$. The final form of 
the electronic susceptibility for a general tight-binding hamiltonian will
be defined by

{\small
$$\frac{1}{2\pi b}\sum_{nm}\alpha_{ij}^{n}\alpha_{ji}^{m}
\biggl(\frac{\sin(m+n+3){\cal A}_{F}}{m+n+3}-\frac{\sin(m+n+1)
{\cal A}_{F}}{m+n+1}\biggr)$$
}

\noindent
which is the final form of the algorithm.
\vspace{5pt}

\section{Discussion on the numerical convergence and tests}

\vspace{10pt}

\hspace{\parindent}To achieve the convergence of the calculation, 
one must ensure first of all, that the total length of the system is much larger that the 
distance between magnetic sites if one wants to consider an infinite medium. To close the system, we take periodic boundary 
conditions. Second, the numerical accuracy must be checked. We found that single precision 
was sufficient and gave the same results as double precision.

\vspace{5pt}

\hspace{\parindent}It remains to figure out how many recursion steps are needed (summation on n
 and m in the above formula). This can be done by noticing that n-recursion steps gives $\frac{n}{W}$ (W the total bandwith)
 as a resolution in energy., and that the states that will mainly contribute to RKKY, are 
 enclosed in an interval of width $\Delta E\leq \hbar/\tau_{L}$ around Fermi energy. Here, we define
 $\tau_{L}$ as the time needed for electrons to travel between two magnetic impurities distant
 of L. For a sufficient resolution of our spectrum, the number of recursion steps to be 
 considered is $n\gg W\tau_{L}/\hbar$.

\vspace{5pt}

\hspace{\parindent}This result may also be recovered by looking at the propagation 
of recursion states in real space. In fact, the propagation of the 
n-th recursion state turns out to be representative of the real wave-function at 
a time $t\sim n\hbar/W$, $\mid \varphi_{i}^{n}\>\sim\mid \Psi(t_{n}\sim n\hbar/W)\>$,
initially located at the same site $\mid r_{i}\>$. Then, one has to check that the diffusion length
of $\mid \varphi_{i}^{n}\>$, which we define as $\xi(n)$, is sufficiently larger than the
distance between magnetic sites. Therefore as soon as $\xi(n)\gg L$, 
the coefficients $\alpha_{ij}^{k>n}$ will not contribute significantly.

\vspace{5pt}

\hspace{\parindent}The criterion $n\gg W\tau_{L}/\hbar$ will depend on the
 nature of propagation through the scattering time. For ballistic regimes (crystals), 
 $\tau_{L}\sim v^{-1}_{F}.L$ with $v_{F}$ the Fermi energy, and $n\gg (W/\hbar v_{F}).L$,
 whereas diffusive regimes will be associated to $L^{2}=D.\tau_{L}$, with $D$ 
 the diffusion constant and $n\gg (W/\hbar D) L^{2}$. Anomalous diffusion described
 by $L\sim A\tau^{\beta}$ ($A,0\leq\beta<0.5$, or $0.5<\beta<1$ depending on the model)
 will lead to $n\gg (W/\hbar A). L^{1/\beta}$ to be control empirically.

\vspace{5pt}

\hspace{\parindent}To test the efficiency of the method, we have checked that the susceptibility
for tight-binding electrons with Fermi energy close to the band edges was equivalent to
 that of free electrons, i.e. well described by the law 
 {\small $\chi (r)\sim (2k_{F}r\sin(2k_{F}r)-\cos(2k_{F}r))/
(k_{F}r)^{3}$} (in 2-dimension). Random systems have also been simulated, by considering site 
 energies distributed at random within $[-W,W]$. For distance larger than 
 the mean free path, the decrease of the averaged susceptibility was found 
 to be proportional to $\frac{1}{r^{2}}\exp(-r/l_{pm})$, and the numerical 
 value of the mean free path was in good agreement with the expected one. 
 Calculations were performed for several values of $W$ with average over 150 configurations.

\section{Applications}

\subsection{2-D Tight-Binding simulation of a Hume Rothery Band Alloy}

\vspace{10pt}

\hspace{\parindent}
In the following, we exemplify the role of local environment on RKKY interaction in a 
2D system. Starting from a perfect crystal, we consider how perturbation of 
local order in the vicinity of the magnetic site may affect RKKY coupling on larger
 (mesoscopic) scales. The local disorder will be designed by a proper
  modification of local tight-binding parameters of the corresponding
   hamiltonian. By doing so, we ensure that the LDoS around the initial
    state $\mid \psi_{0}\rangle$ is strengthened in regards to TDoS at the same energy.

\vspace{5pt}

{\small
\begin{eqnarray}
{\cal H}&=&\sum_{\mid ij\>} \mid ij\>\varepsilon_{ij}\<ij\mid +
 \gamma\sum_{\mid ij\>}
\mid ij\>\bigr(\<ij+1\mid+\<ij-1\mid \nonumber\\
&+&\<i+1j\mid+\<i-1j\mid \bigl)+\tau \sum_{\mid ij\>}
\mid ij\>\bigr(\< \{ i\pm 1 j \pm 1 \} \mid\bigl)\nonumber
\end{eqnarray}
}

\noindent
Our model (Fig.2) features two different hopping integral between first ($\gamma$) and 
second ($\tau$) nearest neighbors as well as an alternate distribution of site 
energies


{\small
$$\epsilon_{ij}={\displaystyle \epsilon_{A}
\over 2} \biggr( 1+(-1)^{i+j}\biggl)+{\displaystyle \epsilon_{B}\over 2}
\biggr(1-(-1)^{i+j}\biggl)$$ 
}

\noindent
in the $\tau =0$ case, this leads to a two-band DOS with a real gap whose bandwith
 $\Delta=\epsilon_{B}-\epsilon_{A}$ ($\epsilon_{A}<\epsilon_{B}$).

\vspace{5pt}

\hspace{\parindent}The orthonormal recursion basis {\small $|\psi_{n}\>$} and 
coefficient can be evaluated iteratively. Nonetheless, even for simple tight-binding 
models, the complexity rapidly increases\cite{Roche}. Taking 
$\varepsilon_{A}=0.1, \varepsilon_{B}=0.3, \gamma=0.1, \tau=0.09$
has TB parameters one gets a TDoS and a RKKY interaction as shown in the inset of Fig.3, 
and one notices in particular that the RKKY coupling shows a structure that
manifests the two underlying periods of the lattice.


\subsection{Effect of local environment on RKKY}

\vspace{10pt}

\hspace{\parindent}Let us now elaborate on the nature of the
 local perturbation around a given magnetic sites. The idea is to 
 particularize the local density of states LDoS around one given site 
 and to measure the effect on the susceptibility on mesoscopic scale.

\vspace{5pt}

\hspace{\parindent}Concretely, we modify the fourth first shells around the initial site
 $\mid \psi_{0}\rangle$ in the following way
$\gamma'=0.65\ \gamma, \tau'=1.26\ \tau$ and the site energies respectively $\varepsilon_{A,B}'\ =\ \varepsilon_{A,B}+\lambda_{i}, \ i=\{1,2,3,4\}$ with $\lambda_{1}=0.15,\ \lambda_{2}=0.12, \ \lambda_{3}=0.09, \ \lambda_{4}=0. 06$. Initial site energy is $\varepsilon_{A}=\<\psi_{0}|{\cal H}|\psi_{0}\>$. 
Thereby, the corresponding LDoS 
($\rho_{i}(E)= -2/\pi \Im m\<r_{i}\mid G(z)\mid r_{i}\>$) is increased if Fermi energy 
is $E_{F}=0.176\ eV$ ($\rho_{i}(E)\geq \rho_{Bulk}(E)$) as shown 
on Fig.4. The number of sites considered for the computation of the susceptibility 
in the numerical was about 250.000, and the number of recursion steps around $\sim 200$.


\vspace{5pt}

\hspace{\parindent}The calculated electronic susceptibility for the 
homogeneous and inhomogeneous cases are plotted on Fig.5
from the results, one clearly sees that such enhancement of LDoS 
leads to an increase of the susceptibility. This effect may be at the origin of
peculiar magnetic properties \cite{Berger} in quasicrystals where atomic order is
known to be complex on mesoscopic scales. Assuming that local
environments of magnetic sites are associated with strong local densities of states
(when compared to the average total density of states), then the RKKY coupling could be
anomalously strong when compared with other metallic disordered phases\cite{RO-RKKY}. 
One notes that such strong fluctuations of LDoS at a given energy from site to site
are also thought to be important at the proximity of a metal-insulator transition\cite{Lerner}.

\vspace{5 pt}

\subsection{Anomalous diffusion and RKKY in quasiperiodic systems}

\vspace{5pt}

\hspace{\parindent}General properties of electronic susceptibility in
quasiperiodic systems are difficult to describe when compared with periodic ones.
Indeed, the simple law obtained for free itinerant electrons may be questionable due to intrinsic 
incommensurability. We show in this section, that the use of recursion method gives here some
unique informations about RKKY in quasiperiodic structures, and further provide a framework for
investigating mesoscopic interaction in aperiodic systems. 
Hereafter we consider a 2D-Fibonacci quasilattice for which we will analyse 
the susceptibility in relation with the spectral properties on larger systems, 
by varying the length of our systems (25.000 sites), 
number of recursion steps (up to 500) and the intensity of quasiperiodic potentials. 
Corresponding spectral structure may be found elsewhere \cite{Fibo2D}. The hamiltonian is defined for a Fibonacci-2D quasilattice 
and written in tight-binding basis as

$$
{\cal H}=\sum_{i}\varepsilon_{i} \mid {\bf r}_{i}\>\<{\bf r}_{i}\mid +
 \gamma\sum_{\<ij\>} \mid{\bf r}_{i} \>\<{\bf r}_{j} \mid
$$


\noindent
for which site energies are given by $\varepsilon_{i}=\varepsilon_{ix}+\varepsilon_{iy}$,
with $\varepsilon_{ix},\varepsilon_{iy}=\pm V_{qp}$ (potential strength) according to a 
Fibonacci sequence. Hopping integrals are set constant for 
simplicity ($\<ij\>$ denotes first neighbors). On Fig.6, the TDoS for a 2D quasiperiodic
 Fibonacci quasilattice, as well as typical signature of incommensurate 
long range order are reported. The strength of the quasiperiodic
potential is $V_{qp}=0.7 (\gamma)$ and the susceptibility is given as a funtion
 of the distance between interacting magnetic-site (in a-units, with a the lattice
spacing). From an analysis as a function of the potential strength, we found
 that no Fermi wavelength can be properly defined and oscillations exhibit resurgences that are
absent from the periodic potential. This is a surprising pattern absent for periodic potentials 
for which unique wave-vector (at Fermi level) and continuous decreasing 
of the coupling is found. Such patterns may however remember the local 
fluctuations found that we found in random systems for a given configuration of
 disorder.  This may thus appear as a common feature between the RKKY in
 quasiperiodic and disordered systems on mesoscopic scales.

\vspace{5pt}

\hspace{\parindent}Incommensurability effects on the period can be 
revealed by analysis of the fluctuations of $\log |\chi (r_{i}-r_{j})|$ as a function 
of $\log (r_{i}-r_{j})$. Indeed, if the $V_{qp}=0.0$ case manifests the oscillating 
behavior $\cos(2k_{F}|r|)$ with unique wave-vector, quasiperiodic potential breaks 
this pattern even for small values such as $V_{qp}=0.05(\gamma)$. On Fig. 7, we compare
 three results obtain at $E_{F}=-1.9\gamma$ and $V_{qp}=0.0, 0.05, 0.1$. Appearance of
 site-dependent incommensurate phase shift is thus illustrated and unveil the 
 action of quasiperiodic potential on electronic coherence ($\log |\chi |$ has been rescaled
 for more clarity)


\vspace{5pt}

\hspace{\parindent}We now consider the long range properties of RKKY in these systems.
To that end, let us return to the relation between electronic 
localization and diffusion modes at the origin of interchange coupling. In the classical 
case (Brownian motion), assuming $p({\bf r},t)$ the probability density of finding a random walker 
(or a classical electron) at ${\bf r}$ after time $t$, 
the anomalous diffusion regime (found in a disordered fractal) is expressed 
by {\small $\< {\bf r}^{2}(t)\>_{dis}=\int {\bf r}^{2} p({\bf r},t) 
d^{D}{\bf r}\sim t^{2\nu}$} \cite{Fractals}. Anomalous regime can also be found in quantum 
diffusion, especially in quasiperiodic systems\cite{RochePRL,RocheJPM} and at the metal-insulator transition
(quantum Hall systems)\cite{Aoki}. This has been connected with multifractal 
properties of eigenstates. Indeed if we consider the spreading of wave-packet constructed from multifractal 
eigenstates, one finds numerically that {\small $\< \hat{\bf r}^{2}\>=\<\Psi(t)|\hat{\bf r}^{2}|\Psi(t)\>=
\int{\bf r}^{2} |\Psi({\bf r},t)|^{2}d^{D}{\bf r}\sim t^{2\nu}$} with $\nu$
 an exponent characterizing the strength of the potential, the intrinsic correlations, 
 etc. According to an argument by E. Akkermans \cite{Akkermans} the power-law decreasing of
 the averaged second moment of electronic susceptibility should not be affected 
by anomalous diffusion and remains of type  $\< \chi(|(r_{i}-r_{j}|)\>
\sim |r_{i}-r_{j}|^{-D}$ for D-dimensional disordered systems.  

\vspace{5pt}

\hspace{\parindent}For the quasiperiodic strengthes considered in our calculation 
(from $V_{qp}=0.05(\gamma)$ up to $V_{qp}=1.1(\gamma)$), the long-range power law 
is not affected qualitatively and one recovers the $1/|r|^{-D}$ with $D=2$. This is 
 exemplified on Fig.8 for two different intensities $V_{qp}=0.1$ and 
$V_{qp}=1.1$ compared with the long ranged oscillations 
for no quasiperiodic potential which mimic the periodic potential. Note that the 
slight departure from power-law in the case $V_{qp}=1.1$ (and for large distance)
 is due to finite-size effects and can be smoothed out by increasing recursion steps.
 
\vspace{5pt}

\hspace{\parindent}One remarks that results on the octagonal quasiperiodic tiling
have been obtained previously \cite{Jag2}, and also seem to indicate a significant 
site dependence of the electronic susceptibility apparently without alteration 
of the general power-law as a function of distance between magnetic sites. 

\vspace{5pt}

\hspace{\parindent}These results are in agreement with the argument we gave in section II
for the general dependence of {\small $\chi(r_{i},r_{j})=(2\rho(E_{F}))/L^{D}
\Re e\sum_{n,m}\Phi(n)\Phi^{*}(m)/(n+m)$} which gives a correct
interpretation of the universal shape of $L^{-D}$ for a metallic system.
But our assumption made on the 
level distribution does not include any subtle correlations in the spectrum, as those found 
at the metal-insulator transition and associated to a specific level spacing 
distribution\cite{MIT-P}. If 
quasiperiodic potentials induces multifractal states, their general level spacing distribution
has been however found to be described by the Gaussian Orthogonal Ensemble (GOE) of the 
Random Matrix theory\cite{Zhong}. This involves a metallic state of the conduction system 
and our calculations are consequently in agreement with other recent results obtained on quasiperiodic
 systems. Manifestations of multifractality of eigenstates may be rather revealed by strong
 fluctuations of local densities of states \cite{Lerner}.

\vspace{5pt}

\hspace{\parindent}To finish with, one notes that unexpected feature has been also revealed 
from careful analysis of site dependent susceptibility (to be published elsewhere). 
For a potential $V_{qp}=0.1$ and Fermi energies $E_{F}=-0.65(\gamma)$ (resp. $E_{F}=-0.575(\gamma)$) situations where the electronic coupling is purely ferromagnetic 
 (resp. antiferromagnetic) were found\cite{Roche}. Such kind of pattern
 unveil unprecedented signature of complicated localization effects unique to 
 quasiperiodic structures. Related phenomena such as Kondo effect 
 in quasiperiodic systems has been discuss in regards to the same localization effects \cite{Benz}.

\section{Conclusion}

\vspace{10pt}

\hspace{\parindent}A method to investigate RKKY interaction in periodic or aperiodic systems has been
 presented. Using real-space schemes, this approach enables in particular to analyse the effect of local inhomogeneities, quasiperiodic potential and by extension in all situations
where usual diagonalization methods may be limited. Some conclusive check 
of the method have been given in comparison to expected behavior and specific 
models have been studied in order to exemplify the interest of our method.
In particular we have shown that even strong regimes of localization induced by
a quasiperiodic potential do not lead to a qualitative departure 
from general power-law dependence, and LDoS fluctuations could affect the intensity of the
interaction.

\vspace{5pt}

\section{\sc Acknowledgments}
 S.R is indebted to Prof. T. Fujiwara from Department of
Applied Physics of Tokyo University for his kind hospitality 
and continuous support. This work has been conducted within 
the GERMA project (Groupe d'Etudes et de Recherches sur les Materiaux Avances).

\vfill\eject

\section{Figures captions}

\vspace{10pt}

\noindent
Fig. 1. Integral contour for the calculation of the coefficient 
${\cal I}_{mn}$ and semi-elliptic density of states $\rho_{ab}(\omega)$
 used in the case of Chebyshev polynomials.

\vspace{20pt}

\noindent
Fig. 2. Schematic representation of the lattice with different tight-binding
parameters.

\vspace{20pt}

\noindent
Fig. 3. Total density of states for the hamiltonian in the perfect Hume-Rothery alloy
and corresponding RKKY coupling (inset).

\vspace{10pt}

\noindent
Fig. 4. Local density of states around the modified local environment
of one magnetic site.

\vspace{10pt}

\noindent
Fig. 5. Comparison of the electronic susceptibility (in a-units, a lattice parameter)
for the homogeneous
 (bold line) and the inhomogeneous cases.

\vspace{10pt}

\noindent
Fig. 6. RKKY coupling for a quasiperiodic system ($V_{qp}=0.7$)
and different Fermi energies. The inset shows the corresponding total density of states.

\vspace{10pt}

\noindent
Fig. 7. Electronic susceptibility 
 as a function of $\log |\chi|$ for different 
 small quasiperiodic potentials ($V_{qp}=0, 0.05 0.01$)

\vspace{10pt}

\noindent
Fig. 8. (a) Electronic susceptibility as a function of $\log |\chi|$ for small and strong 
 quasiperiodic potentials ($V_{qp}=0.1 1.1$) compared to periodic potential. The inset (b) shows
 the local susceptibility for $V_{qp}=0.1$, the same Fermi energy and for several different
 environment between magnetic sites in the quasiperiodic potential.


\vspace{10pt}




\begin{thebibliography}{12345678}

\bibliographystyle{unsrt}


\bibitem{Magnus}
R. Haydock, Solid state Physics vol 35, ed F. Seits, D. Turnbull
and H. Ehrenreich (New York academic) 216 (1980). 
{\it The recursion method and its applications}
Editors D.G. Petitfor and D.L. Weaire (Springer Verlag),
 Springer Series in Solid State Sciences {\bf 58} (1984).  

\bibitem{RKKY}
M.A. Ruderman and C. Kittel, Phys Rev. {\bf 96}, 99 (1954).
T.  Kasuya, Phys. Rev. {\bf 106}, 893 (1957). 
K. Yosida, Prog. Theor. Phys. {\bf 16}, 45 (1956).

\bibitem{Degennes}
 P.G. De Gennes, {\it J. Phys. Radium} {\bf 23}, 630 (1962).
G. Bergmann,  {\it Phys. Rev. B} {\bf 36}, 2469 (1987).
L.N. Bulaevskii and S.V. Panyukov, {\it JETP Letter} {\bf 43}, 240 (1986).

\bibitem{Jag}
 A. Jagannathan, E. Abrahams and M.J Stephen, {\it Phys. Rev. B}, 37 
(1988) 436. 

\bibitem{Lerner}
I.V. Lerner, {\it Europhys. Lett.}, {\bf 16} (5), 479-484 (1991).

\bibitem{Spinglass}
F. Matsubara and M. Iguchi, {\it Phys. Rev. Lett.} {\bf 68}, 3781 (1992)

\bibitem{HTC}
J.J. Rodriguez-Nu$\tilde{n}$ez, H. Beck, J. Konior, A. M. Oles and B. Coqblin,
{\it Phys. Lett. A} {\bf  197}, 173 (1995).


\bibitem{GMR}
P. Bruno and C. Chappert, {\it Phys. Rev. Lett.} {\bf 67}, 1602 (1991).
B.A. Jones and C.B. Hanna,  {\it Phys. Rev. Lett.} {\bf 71}, 4253 (1993).
P. Bruno, J. Kudrnovsky, V. Drchal and I. Turek, {\it Phys. Rev. Lett.} 
{\bf 76}, 4254 (1996).


\bibitem{RochePRL}
S. Roche and D. Mayou, {\it Phys. Rev. Lett.} {\bf 79}, 2518 (1997).

\bibitem{RocheJPM}
S. Roche, G. Trambly de Laissardiere and D. Mayou, 
{\it J. Math. Phys.} {\bf 38}, 1794 (1997).

\bibitem{Berger}
C. Berger and J.J. Pr\'ejean, {\it Phys. Rev. Lett.} {\bf 64}, 1769 (1990);
A. Gozlan et al, {\it Phys. Rev. B} {\bf 44}, 2 (1991).

\bibitem{Roche}
S. Roche, Ph.D. Thesis, University Joseph-Fourier (1996), unpublished.


\bibitem{RO-RKKY}
S. Roche and D. Mayou, {\sl Proc.
of the 5th International Conference on Quasicrystals, Avignon, 1995}, Editors C. Janot, 
R. Mosseri, 413 (World Scientific, 1995).


\bibitem{Szego}
G. Szeg\"o: {\it Orthogonal polynomials} 4th ed. Colloquium
Publications {\bf 23} (Providence, Rhode Island: Am. Math. Soc. 1975).

\bibitem{Fractals}
S. Havlin and D. Ben Avraham, {\it Adv. Phys.} {\bf 36}, 695 (1987).
J.P. Bouchaud and A. Georges, {\it  Physics Reports} {\bf 195}, 131 (1990)

\bibitem{Aoki}
B. Huckestein and L. Schweitzer, {\it Phys. A} {\bf 191}, 406 (1991). 
Takamichi Terao, Tsuneyoshi Nakayama and H. Aoki, 
{\it Phys. Rev. B} {\bf 54}, 10350 (1996).


\bibitem{Akkermans}
E. Akkermans (unpublished work).

\bibitem{Fibo2D}
X. Fu, Y. Liu, B. Cheng and D. Zheng, {\it Phys. Rev. B} {\bf 43}, 10808 (1991).

\bibitem{Jag2}
A. Jagannathan, {\it J. Phys. I (France)} {\bf 4}, 133 (1994).

\bibitem{MIT-P}
I.K. Zharekeshev and B. Kramer, {\it Phys. Rev. Lett.} {\bf 79}, 717 (1997).

\bibitem{Zhong}
J.X. Zhong, U. Grimm. R.A. Romer and M. Schreiber, {\it Phys. Rev. Lett.} {\bf 80}, 
3996 (1998).


\bibitem{Benz}
V.G. Benza and E. Montaldi ,{\it J. Phys. A: Math. Gen.} 
{\bf 27}, 2299-2304 (1994).


\end{thebibliography}
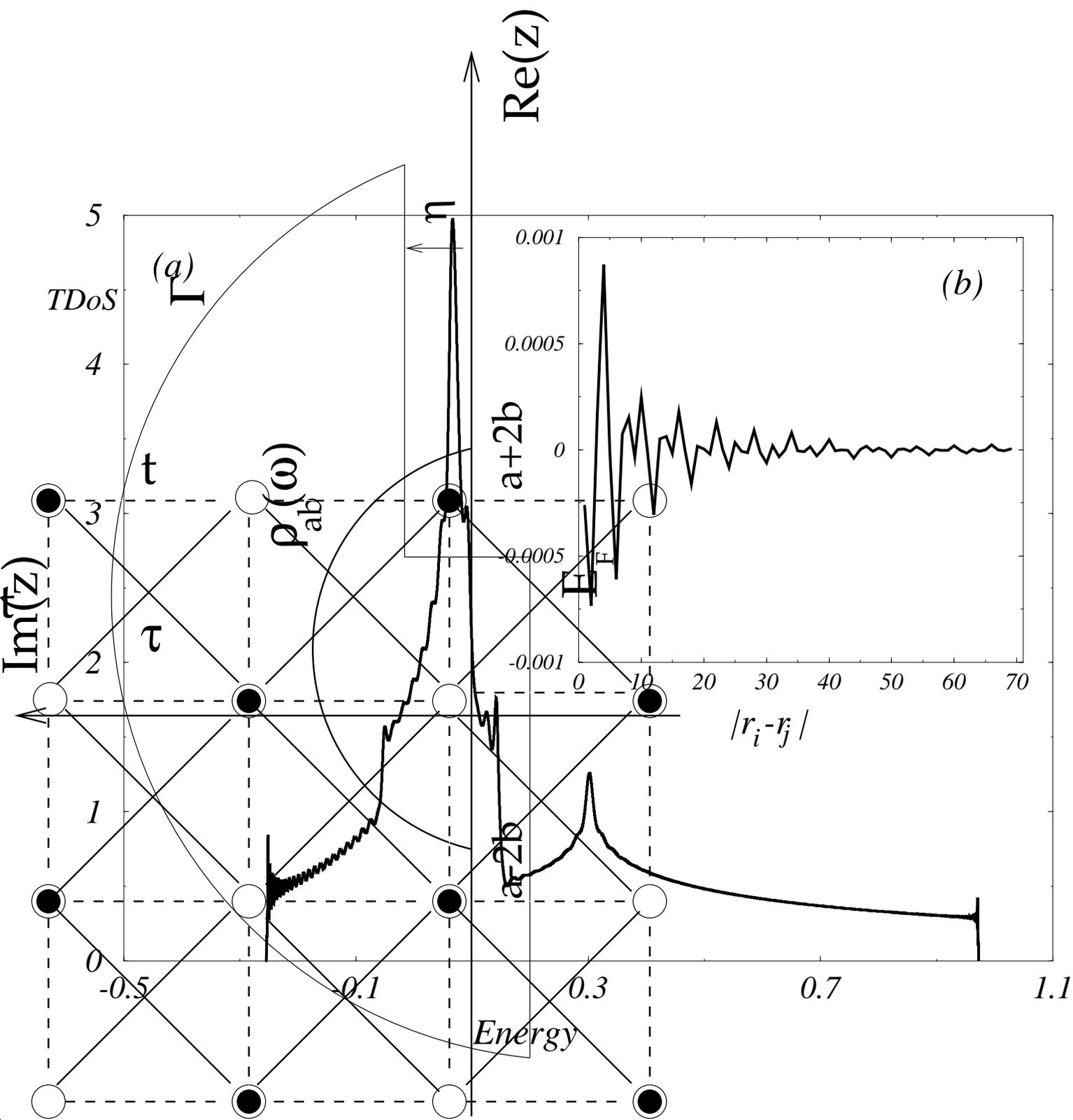
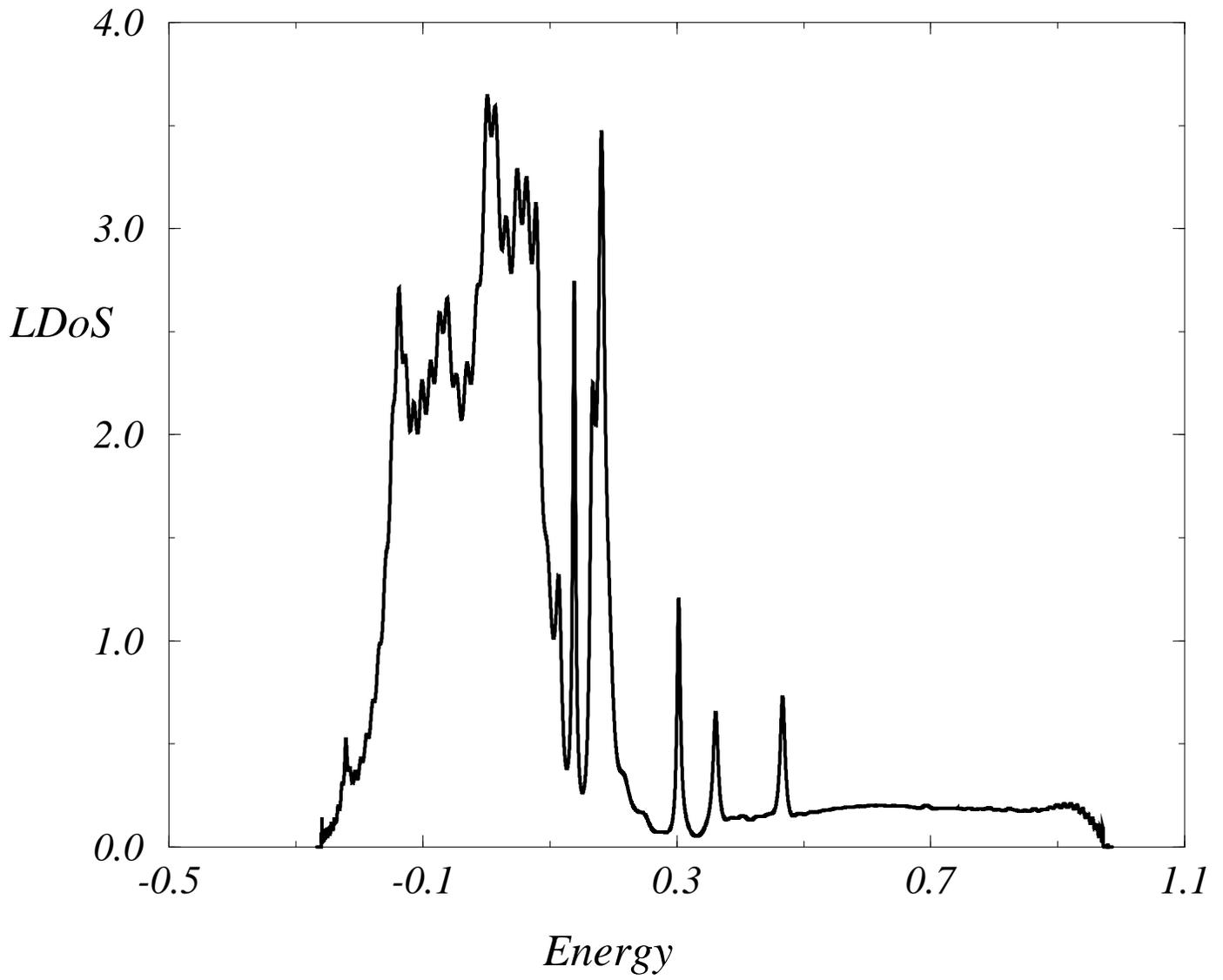
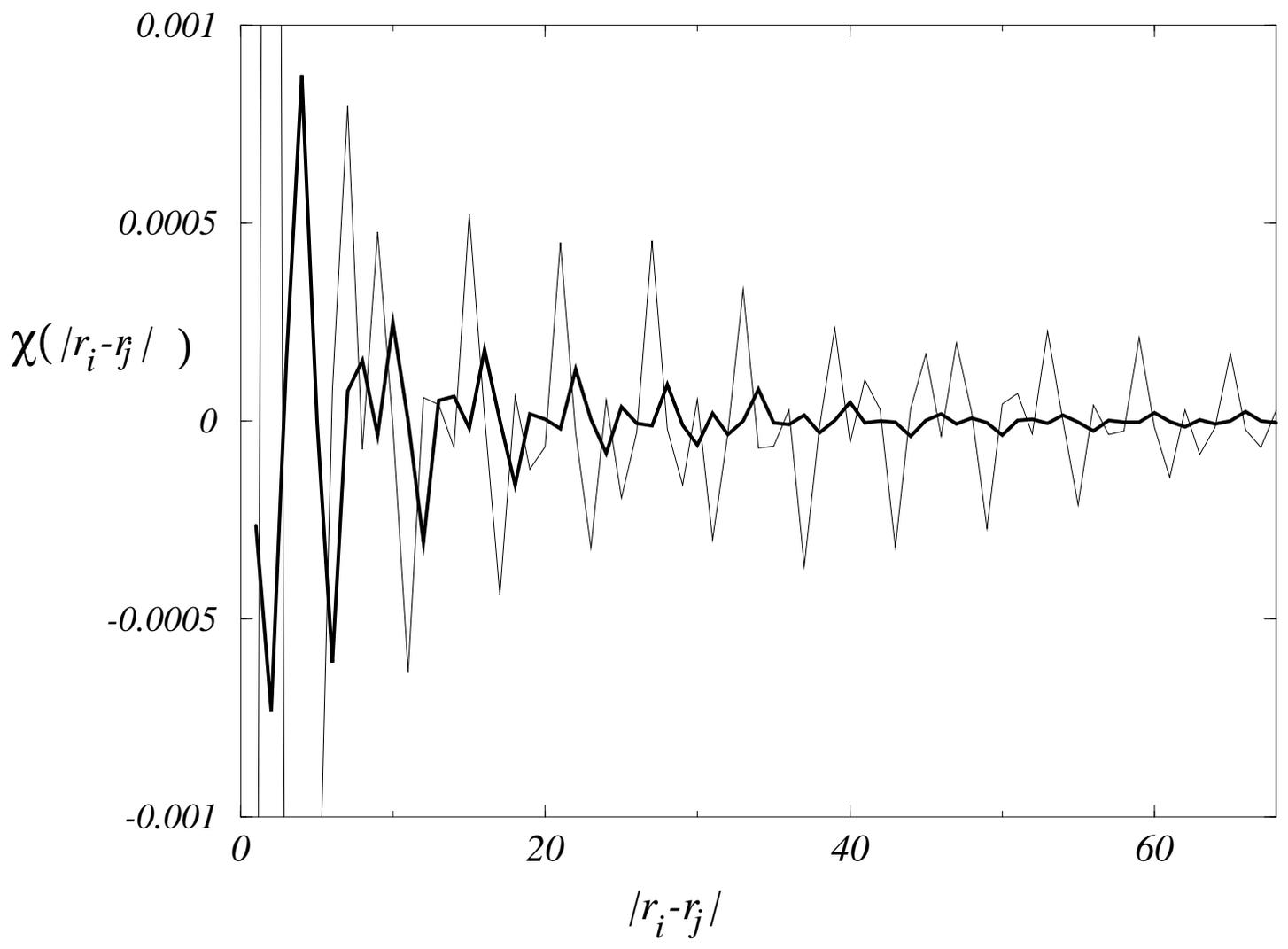
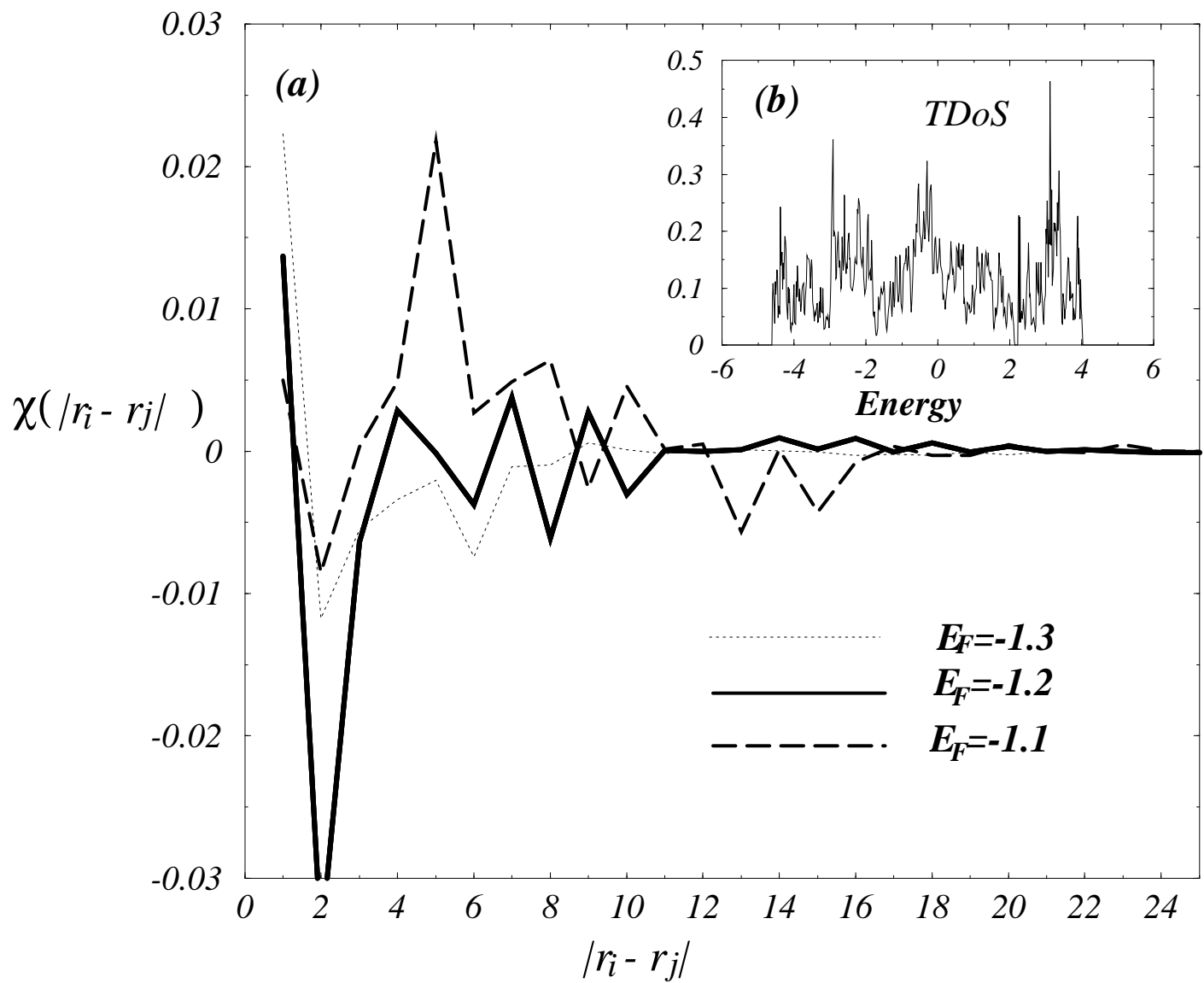
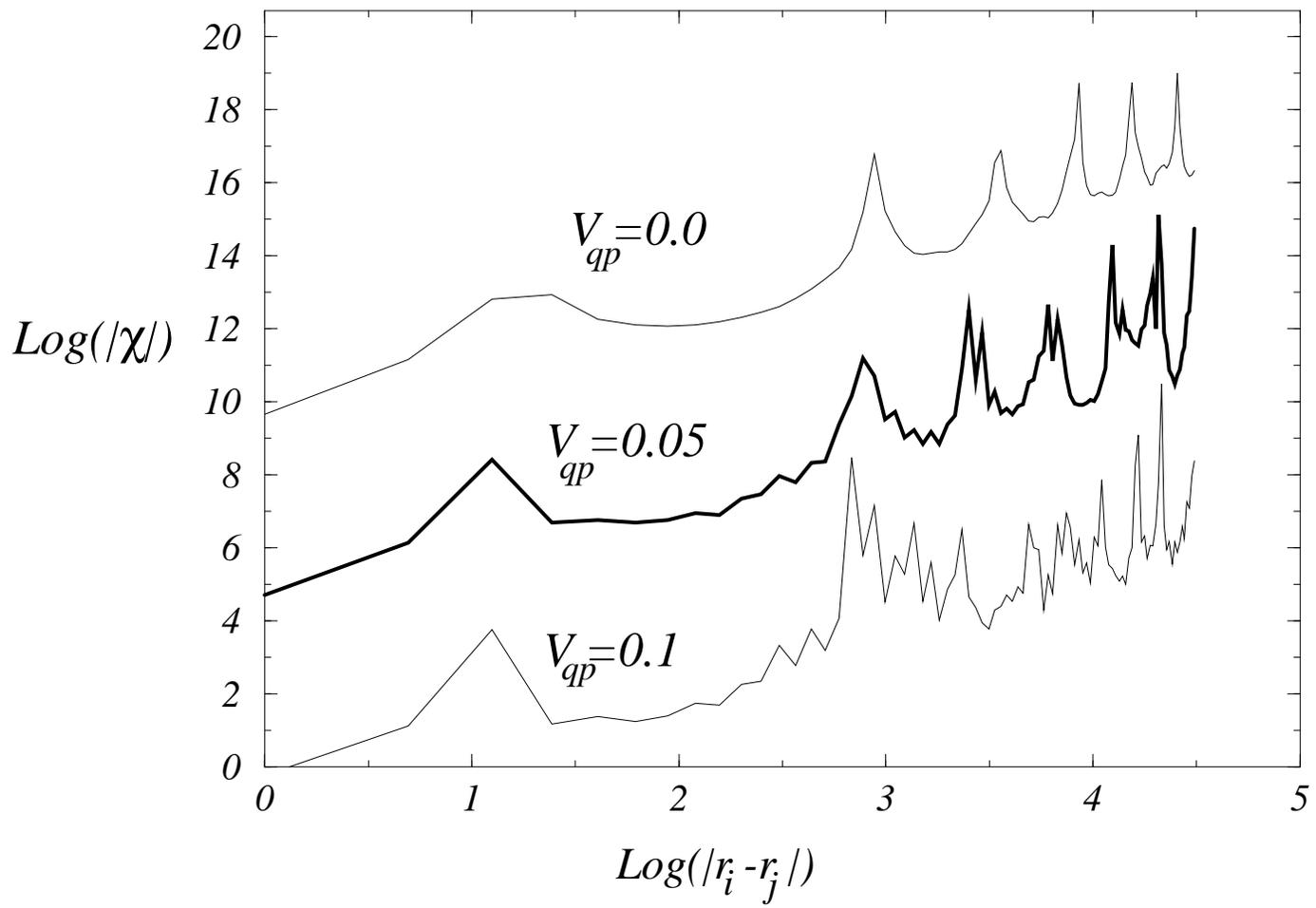
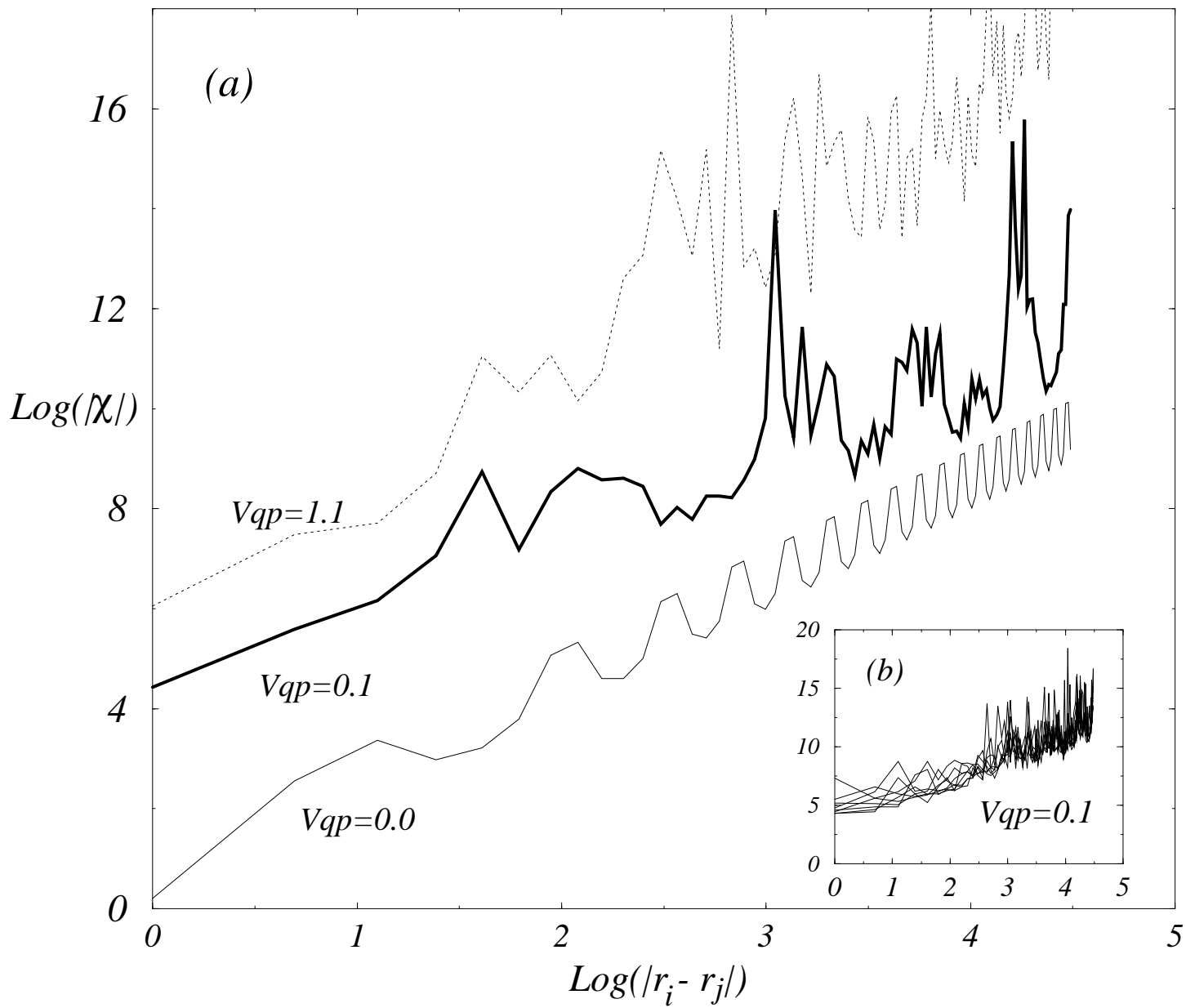
\end{document}